# Accurate relative-phase and time-delay maps all over the emission cone of hyperentangled photon source

Salem F. Hegazy[a,b], Jala El-Azab [a], Yehia A. Badr[a], Salah S. A. Obayya[b I]

[a] National Institute of Laser Enhanced Sciences, Cairo University, 12613 Giza, Egypt
[b] Centre for Photonics and Smart Materials, Zewail City of Science and Technology, 12588 Giza, Egypt

## ABSTRACT

High flux of hyperentangled photons entails collecting the two-photon emission over relatively wide extent in frequency and transverse space within which the photon pairs are simultaneously entangled in multiple degrees of freedom. In this paper, we present a numerical approach to determining the spatial-spectral relative-phase and time-delay maps of hyperentangled photons all over the spontaneous parametric down conversion (SPDC) emission cone. We consider the hyperentangled-photons produced by superimposing noncollinear SPDC emissions of two crossed and coherently-pumped nonlinear crystals. We adopt a vectorial representation for all parameters of concern. This enables us to study special settings such as the self-compensation via oblique pump incidence. While rigorous quantum treatment of SPDC emission requires Gaussian state representation, in low-gain regime (like the case of the study), it is well approximated to the first order to superposition of vacuum and two-photon states. The relative phase and time-delay maps are then calculated between the two-photon wavepackets created along symmetrical locations of the crystals. Assuming monochromatic plane-wave pump field, the mutual signal-idler relations like energy conservation and transverse-momentum conservation define well one of the two-photon with reference to its conjugate. The weaker conservation of longitudinal momentum (due to relatively thin crystals) allows two-photon emission directions coplanar with the pump beam while spreading around the perfect phase-matching direction. While prior works often adopt first-order approximation, it is shown that the relative-phase map is a very well approximated to a quadratic function in the polar angle of the two-photon emission while negligibly varying with the azimuthal angle.

## 1. INTRODUCTION

Simultaneous entanglement of multiple degrees of freedom, or so called "hyperentanglement", has been widely used to circumvent limitations of linear[1] and nonlinear[2] optics in performing several quantum operations like Bell state analysis[3-5], superdense coding[6], and superdense teleportation[7]. Another vital advantage emerges because the quantum logic between qubits in different degrees of freedom of one photon is relatively easy when compared to qubits based in different photons[8,9]. The noncollinear spontaneous parametric down conversion (SPDC) emission of two coherently-pumped, crossed, and adjacent nonlinear crystals has been widely used to generate hyperentangled photons[10-18]. There have been several attempts[11-18] to idealize the output of the two-crystal source by erasing the which-crystal information, e.g., enhancing the temporal overlap of the down-converted photons[11], and by flattening the angle-dependent relative-phase[12].

However, today, 17 years after its invention, the high-fidelity emission of the hyperentangled photons source is still limited to relatively small apertures, owing to the angle-dependent relative-phase. There are several simplifications that have been adopted to determine the relative phase of the two-photon state produced by the two-crystal emission. In their way to demonstrate the brightest -at that time- high-fidelity source of polarization entangled photons, Altepeter *et al.*[12] presented the first attempt to calculate the spatial relative-phase variation which was restricted to SPDC light within the horizontal $xz$ plane. It was important to define and compensate for the relative-phase variations so as to enable collection over wide angles while avoiding the well-known flux-purity tradeoff. Their calculations came along with experimental results via quantum state tomography over 25 points spreading all over the transverse detection window. However, the measurements clarified a significant mismatch with the theory which was referred to the possibility of unjustified approximations with the calculations. After two years, an erratum[13] was published to correct the external relative-phase term which apparently increased the matching with the experimental results. The revised method was subsequently used by Rangarajan *et al.*[14] to run simultaneous compensation for spatial relative-phase variations and time delay between the coherent emissions of the two crystals.

[I] sobayya@zewailcity.edu.eg

However, all of these works[12-14] consider the phase as a function in space only (i.e., for monochromatic SPDC), while the phase also varies in frequency (which gains more importance when wide spectral filters are in use). Not only but they also double the phase accumulated by one of the two-photon to obtain the relative-phase of the two-photon state which was then verified invalid[18] in their case[II].

An elegant method for relative-phase calculation was then presented by Cialdi *et al.*[16] by approximating the phase to first order in a number of spatial and spectral parameters. However, the accuracy of this approach is limited to small ranges of emission angles and spectrum where the map linearity remains a good approximation.

Recently, we determined the exact relative-phase as a function of frequency and angle of emission, taking into consideration the finite spectral width of the pump beam[18]. To illustrate the spatial-spectral phase map in this case, we first considered that the pump beam is monochromatic plane wave and the SPDC light emission is in directions not restricted to the perfect phase matching condition. This reduced the parameters with which the phase map varies to the frequency and the emission angle of the signal photon. The compensation for this sort of phase variations can be then achieved by two birefringent elements inserted within the collection angles of downconverted photons. The tunability of the compensated phase function was offered by tilting the two birefringent elements.

Second, we considered the contrast situation: the pump beam is of a finite line width and the SPDC light is restricted to perfect phase matching directions. This alternatively reduces the phase map parameters to the frequencies of signal and idler photons. The variations of the phase map come then only with the sum of the downconversion frequencies; that is the frequency of the pump beam. To compensate for this spectral phase gradient, we offered two approaches, either to use different parameters of the two birefringent elements acting on the SPDC light (which is found a complex problem and may be unrealizable with the available birefringent materials), or to insert a third birefringent element to the path of the pump beam before striking the two crystals. The later approach, besides being easily realizable, offered the important result that the compensation for the phase gradient due to pump spectrum and under the perfect phase matching condition is equivalent to the compensation needed for the delay between the two-photon wavepackets produced in the first and the second crystals. This approach while enables the accurate calculation of the relative-phase over wide ranges of spectrum and emission angle, it is –like previous approaches– limited to SPDC light within the horizontal $xz$ plane.

However, the detection of high-fidelity polarization entangled photons over ultra-wide apertures requires compensation for the angle-dependent relative-phase over all the spatial acceptance range, and therefore, necessitates determining the relative-phase map all over the SPDC cone. On the other hand, to verify the purity of the emission, other properties like the time delay between the two-photon wavepackets emerging from the 1st and 2nd crystals need to be determined all over the emission cone.

In this paper, we adopt a vectorial representation to describe all the phase-matching parameters such as the wavevectors, the Poynting vectors of the ordinary and the extraordinary components of pump and SPDC light traversing the two crystals. This enables us to determine the maps of the relative phase and time delay over the 2D angular space (defined by the polar and azimuthal angles of SPDC emission). The vectorial representation enables also determining the relative-phase and time-delay maps in special settings such as the oblique incidence of the pump beam which can be used for self-compensation.

---

[II] For negative birefringent crystals such as BBO and $LiIO_3$, the relative phase of the produced two-photon state can be as double as the phase accumulated by one of the two-photon if the principal plane of the 2nd crystal is orthogonal to the plane where the relative phase is calculated.

## 2. THE TWO-PHOTON STATE OF THE SUPERIMPOSED DOWN-CONVERSIONS

Consider two orthogonal type-I crystals coherently pumped by diagonally polarized pump beam. The two crystals are of negative birefringence and infinite lateral extent. The two-photon state of the superimposed SPDC emissions is entangled in polarization, spatial mode, and energy-time. The polarization part of the two-photon state can be expressed as

$$|\Psi\rangle = \frac{d}{\sqrt{2}}\int d^2q_s\, dw_s\, \chi_{eff}^{(2)}(w_s, w_p - w_s; w_p)\, \text{sinc}\left[\frac{\Delta\kappa d}{2}\right] \exp\left[i\frac{\Delta\kappa d}{2}\right]$$
$$\times \left\{|H_{q_s,w_s}\rangle|H_{q_i,w_i}\rangle + \exp[i\phi_o + i\phi_{DC}(q_s,w_s)]\,|V_{q_s,w_s}\rangle|V_{q_i,w_i}\rangle\right\} \qquad (1)$$

where $w_p$ and $w_s$ are the angular frequencies of pump and signal photons, respectively, $\chi_{eff}^{(2)}$ is the effective susceptibility, $q_s$ is the transverse wavevector component of the signal, $d$ is the crystal length, $\Delta\kappa$ is the longitudinal wavevector mismatch, $H$ and $V$ denote the horizontal and vertical polarizations, $\phi_o$ is the phase difference initially existing between orthogonal pump components, and $\phi_{DC}(q_s, w_s)$ is the relative phase associated with the downconversion process. In (1), the pump beam is assumed to be a monochromatic plane wave; so that the conservation of energy and transverse momentum reduces the integrands to be on signal parameters only. It is worth mentioning that if the biphoton is restricted to single-mode detection, the spatial dependence of $\phi_{DC}$ effectively vanishes, leaving solely the spectral dependence.

## 3. VECTORIAL REPRESENTATION

### 3.1 Normal pump-beam incidence

It is straightforward to parameterize the two-crystal emission in the 2D angular space $(\theta, \phi)$ by adopting the vectorial analysis instead of the conventional scalar analysis in the case of 1D space[18]. We represent the optic-axes of the 1st and 2nd crystals as the unit vectors

$$a_n = (\sin\theta_{a_n}\cos\phi_{a_n}, \sin\theta_{a_n}\sin\phi_{a_n}, \cos\theta_{a_n}),\ \ n = 1,2 \qquad (2)$$

where $\theta_{a_n}, \phi_{a_n}$ are (polar and azimuthal) angles of the optic axis of crystal $n$ as shown in Fig. 1. As depicted in Fig. 1, the unit wavevectors of the pump photon (taken to be collinear with z axis) and the scattered ordinary signal and idler photons are

$$K_{pe} = (0,0,1), \qquad K_{\varepsilon o} = (\sin\theta_{\varepsilon o}\cos\phi_{\varepsilon o}, \sin\theta_{\varepsilon o}\sin\phi_{\varepsilon o}, \cos\theta_{\varepsilon o}), \qquad \varepsilon = s, i \qquad (3)$$

where the angles of $K_{so}$ and $K_{io}$ are related together via the transverse momentum conservation. After crossing the inter-crystal interface, ordinary downconverted light turns to be extraordinarily polarized with the corresponding unit wavevectors

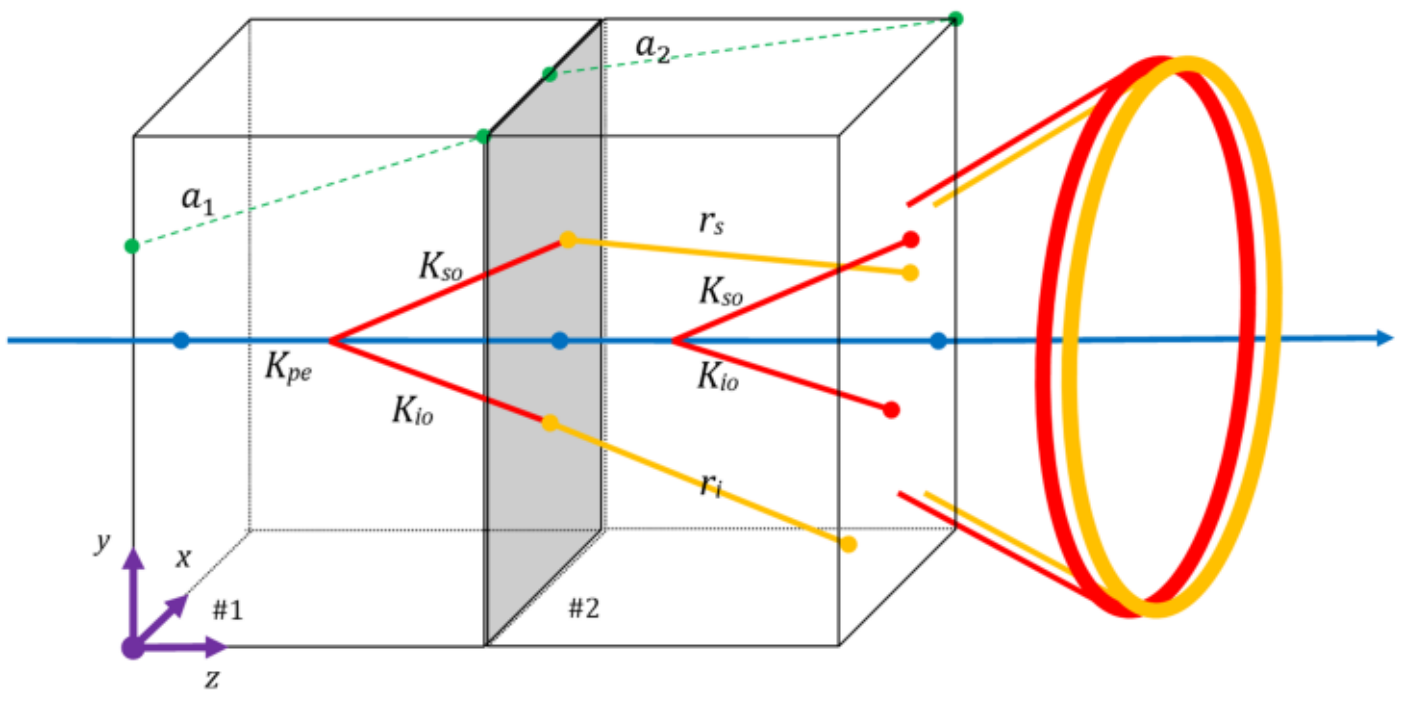

Figure 1. Two crystal source in the case of normal pump incidence

$$K_{\varepsilon e} = (\sin\theta_{\varepsilon e}\cos\phi_{\varepsilon o}, \sin\theta_{\varepsilon e}\sin\phi_{\varepsilon o}, \cos\theta_{\varepsilon e}), \qquad \varepsilon = s, i \tag{4}$$

Here while the azimuthal angles (the $\phi$'s) are strictly the same as $K_{so}$ and $K_{io}$ (the rays refracted into the 2nd crystal lie within the plane of incidence), the polar angles (the $\theta$'s) are different due to the different refractive indices. Because the extraordinary refractive index itself is angle dependent, the polar angles cannot be given in a direct way. Several iterations should be made to obtain precisely the value of $\theta_{\varepsilon e}$ using the refraction relation

$$\sin^2\theta_{\varepsilon e} = n_o^2(w_\varepsilon)\sin^2\theta_{\varepsilon o}\left[\frac{(a_2 \,.\, K_{\varepsilon e})^2}{n_o^2(w_\varepsilon)} + \frac{1-(a_2 \,.\, K_{\varepsilon e})^2}{n_e^2(w_\varepsilon)}\right], \qquad \varepsilon = s, i \tag{5}$$

where $n_o$ and $n_e$ are the principal values of the refractive index. These iterations should be carried out all over the 2D space of the polar and azimuthal angles. The corresponding extraordinary wavevectors can be then determined by multiplying the unit vector by the wavenumber which is function of the frequency and the angle

$$\alpha_\varepsilon = \arccos(K_{\varepsilon e}.\, a_2), \quad \varepsilon = s, i, \tag{6}$$

that is variable across the 2D space. In the 2nd crystal, the Poynting unit vectors $r_s$, $r_i$ can be obtained by substituting the walk-off angle

$$\sigma_\varepsilon = \left[\alpha_\varepsilon - \operatorname{atan}\left(\frac{n_o^2(w_\varepsilon)}{n_e^2(w_\varepsilon)}\tan\alpha_\varepsilon\right)\right]\operatorname{sign}[n_o(w_\varepsilon) - n_e(w_\varepsilon)] \tag{7}$$

into the decomposition relation

$$r_\varepsilon = K_{\varepsilon e}\cos\sigma_\varepsilon + \operatorname{sign}[n_o(w_\varepsilon) - n_e(w_\varepsilon)]\,\frac{a_2 - [K_{\varepsilon e}.\, a_2]K_{\varepsilon e}}{|a_2 - [K_{\varepsilon e}.\, a_2]K_{\varepsilon e}|}\sin\sigma_\varepsilon, \qquad \varepsilon = s, i \tag{8}$$

This form positions the Poynting vector $r_\varepsilon$ within the plane of $K_{\varepsilon e}$ and $a_2$ at an angle $\sigma_\varepsilon$ off the wavevector $K_{\varepsilon e}$. Finally, the external (free-space) unit wavevectors can be calculated as

$$K_{\varepsilon a} = \left(n_o(w_\varepsilon)\sin\theta_{\varepsilon o}\cos\phi_{\varepsilon o}, n_o(w_\varepsilon)\sin\theta_{\varepsilon o}\sin\phi_{\varepsilon o}, \sqrt{1 - n_o^2(w_\varepsilon)\sin^2\theta_{\varepsilon o}}\right), \qquad \varepsilon = s, i \tag{9}$$

### 3.2 Oblique pump-beam incidence

The key idea in the calculations of the oblique-incidence situation is to consider the initial unit vectors $K_{so}$ and $K_{io}$ with respect to the pump frame $(x', y', z')$ defined as in Fig. 2. The pump frame has $z$-axis collinear to the interacting component of the pump beam (extraordinary component in case of negative birefringence crystal) and $x$-axis and $y$-axis given by the polar and azimuthal angles ($\theta_p$ $and$ $\phi_p$) of the interacting pump component (internal to the crystal). The pump frame is therefore aligned to the laboratory frame in case of normal pump incidence with $\theta_p = 0$ and $\phi_p = 0$. The axes of the pump frame are related to the laboratory frame as

$$\begin{bmatrix} x \\ y \\ z \end{bmatrix}_{laboratory} = \begin{bmatrix} \cos\theta_p \cos\ \phi_p & -\sin\phi_p & \sin\theta_p \cos\phi_p \\ \cos\theta_p \cos\phi_p & \cos\phi_p & \sin\theta_p \sin\phi_p \\ -\sin\theta_p & 0 & \cos\theta_p \end{bmatrix} \begin{bmatrix} x' \\ y' \\ z' \end{bmatrix}_{pump} \quad (10)$$

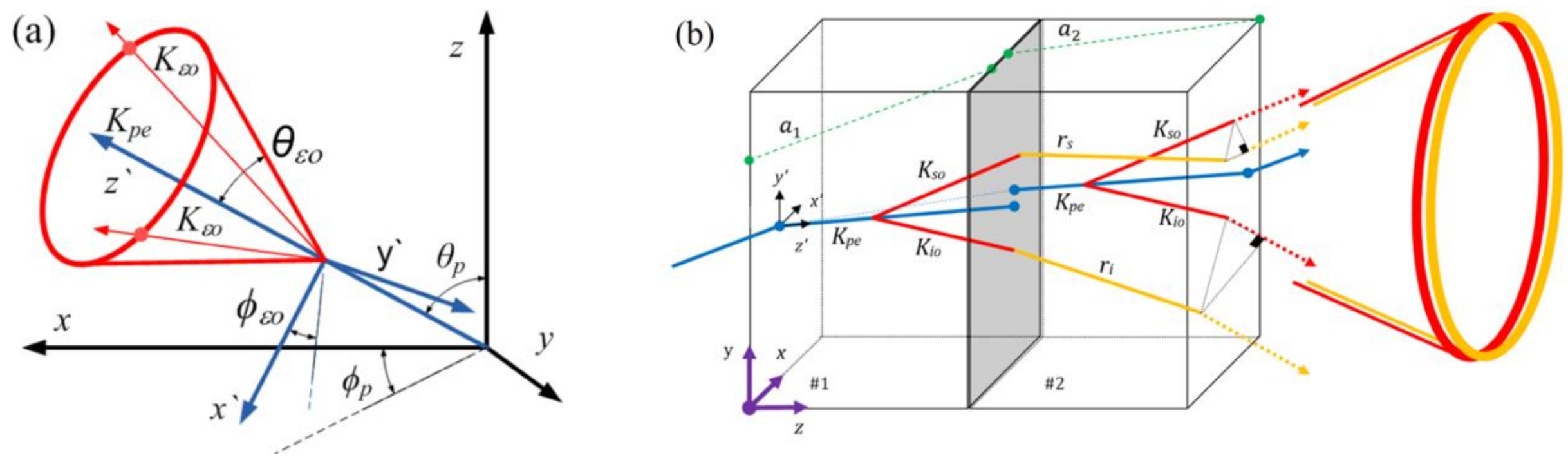

Figure 2. (a) Laboratory frame $(x, y, z)$ and pump frame $(x', y', z')$. (b) Two crystal source in the case of oblique pump incidence.

As in Fig. 2, it is straightforward to describe SPDC emission all over the cone (and even the optic axes) with reference to the pump frame. Using transformation of axes, we can redefine all vectors in the laboratory frame.

It is worth noticing that the phase matching angle (the angle between the interacting component of the pump beam and the optic axis) should be kept the same along the traverse of the pump beam through the two crystals, thereby maintaining the overlap of the superimposed SPDC light.

## 4. MAPS ALL OVER THE TWO-CRYSTAL EMISSION

In this section, we determine the spatial and spectral maps of the relative-phase and time-delay that can be observed via multimode two-photon detection.

### 4.1 Relative-phase map

The relative-phase acquired by the two-photon state in (1) can be divided into four parts

(A) The phase accumulated due to the pass of the 1st-crystal emission across the 2nd crystal.
(B) The phase difference between the emissions of the two crystals as being diffused in free space from different points of origin.
(C) The relative phase due to the different positions of the two crystals along the $z$ axis.
(D) The phase difference initially exists between orthogonal pump components.

The phase map can be calculated all over the two-photon emission cone as

$$\phi_{DC}(w_s, \theta_{sa}, \phi_{sa}) = \sum_{\varepsilon=s,i} \frac{w_\varepsilon\ d}{c\ r_{\varepsilon,z}} \left[n_e(w_\varepsilon, \alpha_\varepsilon)\ \ (K_{\varepsilon e}.r_\varepsilon) + (r_{\varepsilon,x}\ , r_{\varepsilon,y}, 0).K_{\varepsilon a}\right] \quad (11)$$

which is function of the signal parameters $(w_s, \theta_{sa}, \phi_{sa})$ only, thanks to the one-to-one mapping to the idler's $(w_i, \theta_{ia}, \phi_{ia})$. It is therefore sufficient to plot the relative phase map $\phi_{DC}(w_s, \theta_{sa}, \phi_{sa})$ over half the 2D angular space (the half corresponding to the signal photon) as shown in Figs. 3-6. In these figures, one can observe no significant difference in the relative-phase map for different settings of pump's angle-of-incidence. It worth mentioning that within the examined angular ranges, the relative-phase function all over the cone is very well approximated to a quadratic function in the polar (emission) angle while slightly varying with the azimuthal angle of the two-photon. This approximation is verified by the linear slopes of the phase maps.

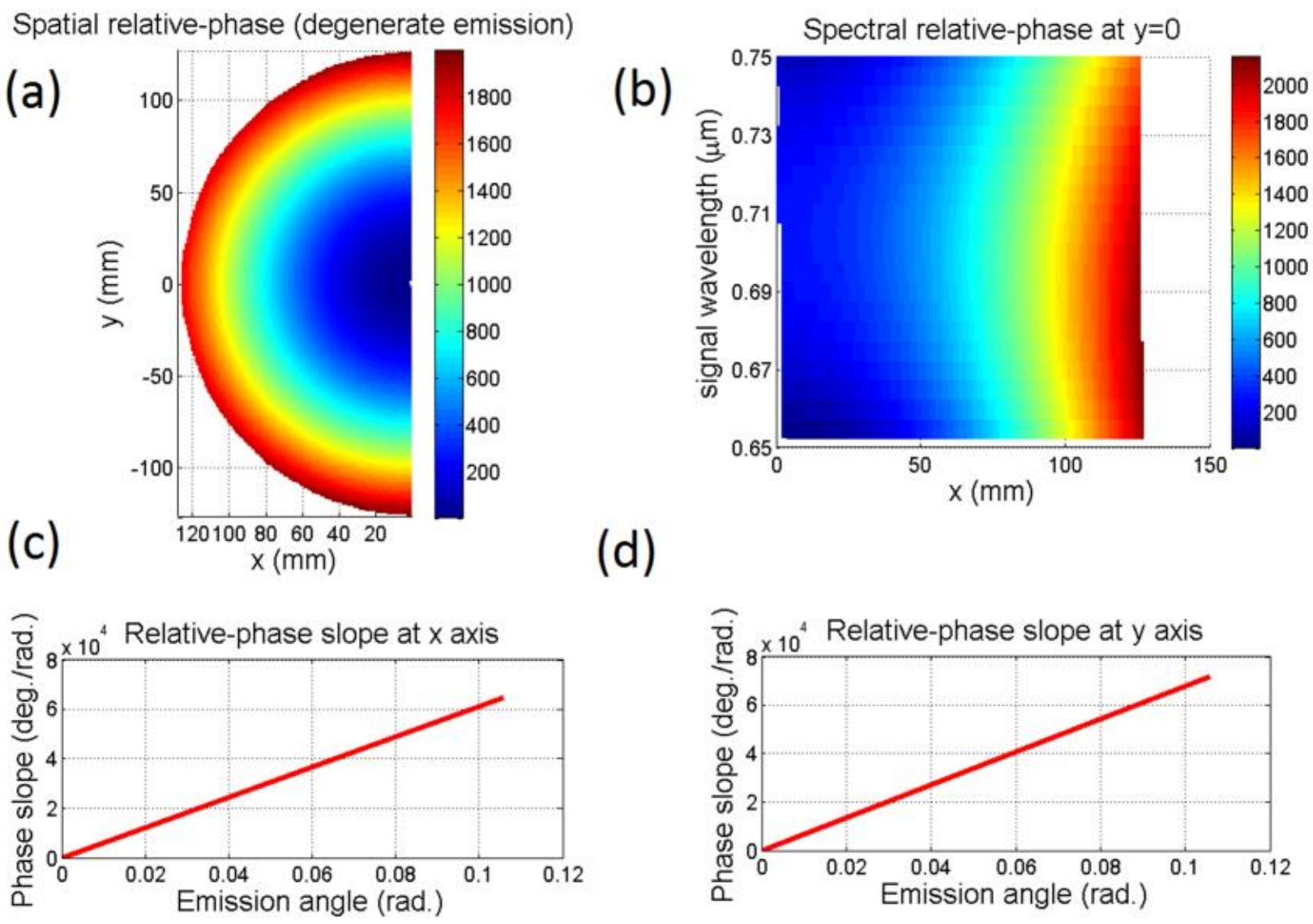

Figure 3. (a) The spatial relative-phase map (in degrees) for the degenerate emission of two $LiIO_3$ crystals of 0.59 mm length with the phase-matching angle between the pump beam and optic axis equals 51.95° The two-crystal are pumped by normally incident beam ($\theta_p = 0°$ and $\phi_p = 0°$) at 351.1-nm to produce degenerate emission at ~3°. The phase maps are calculated with reference to the position of signal photon (occupying the space $\phi = [-90°, 90°]$) at a transverse plane, 120 cm from the two crystals. The axes $x$ and $y$ are those of the Laboratory frame (therefore, fixed for oblique incidence) where the origin point represents the point that the pump beam passes in case of normal incidence. (b) The spectral profile for the relative phase along $y = 0$ line for SPDC emission around the degenerate frequency. (c) and (d) The angular slope of the relative phase along $y = 0$ and $x = 0$, respectively.

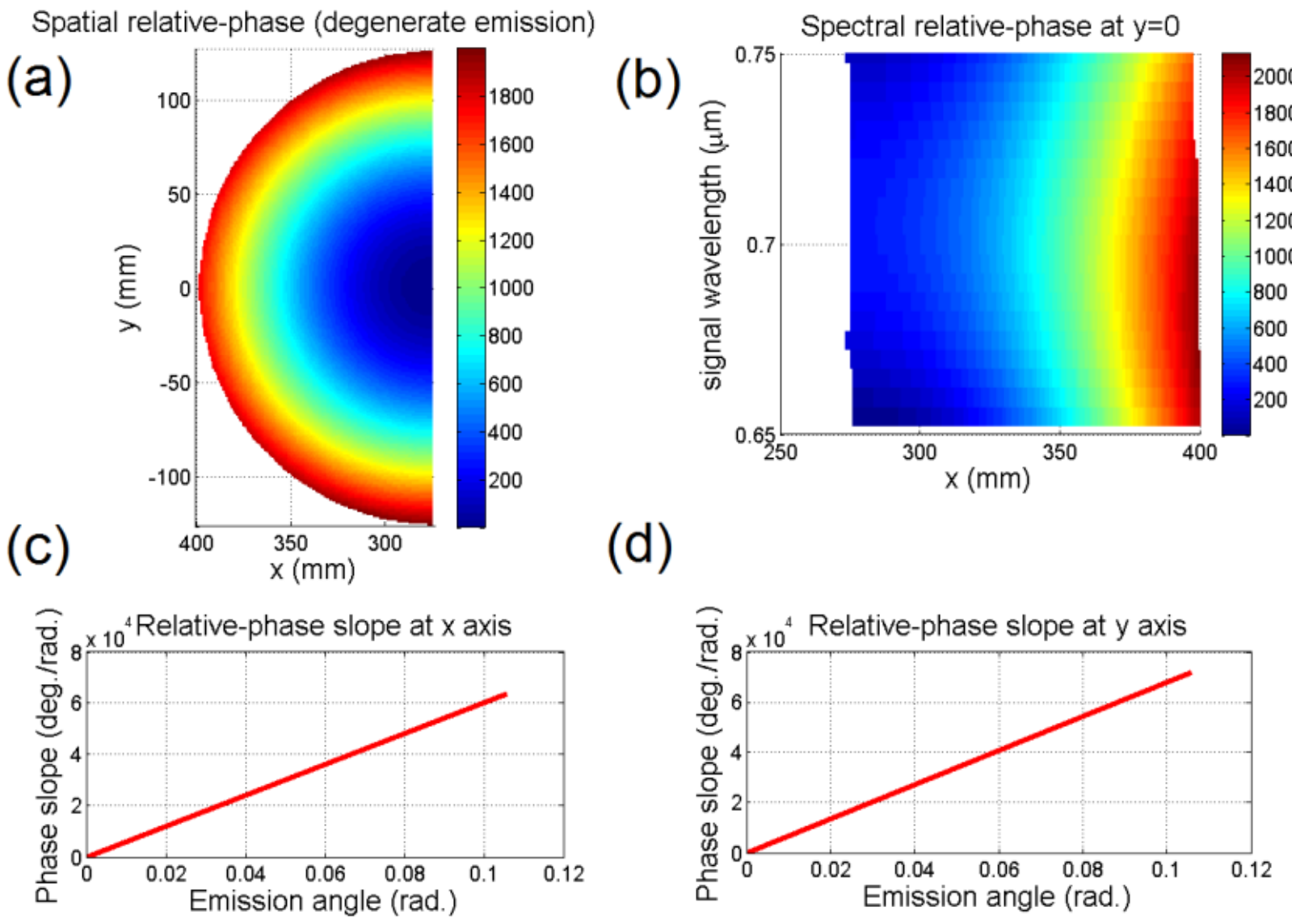

Figure 4. The same as in Fig. 3 for $\theta_p = 7°$ and $\phi_p = 0°$

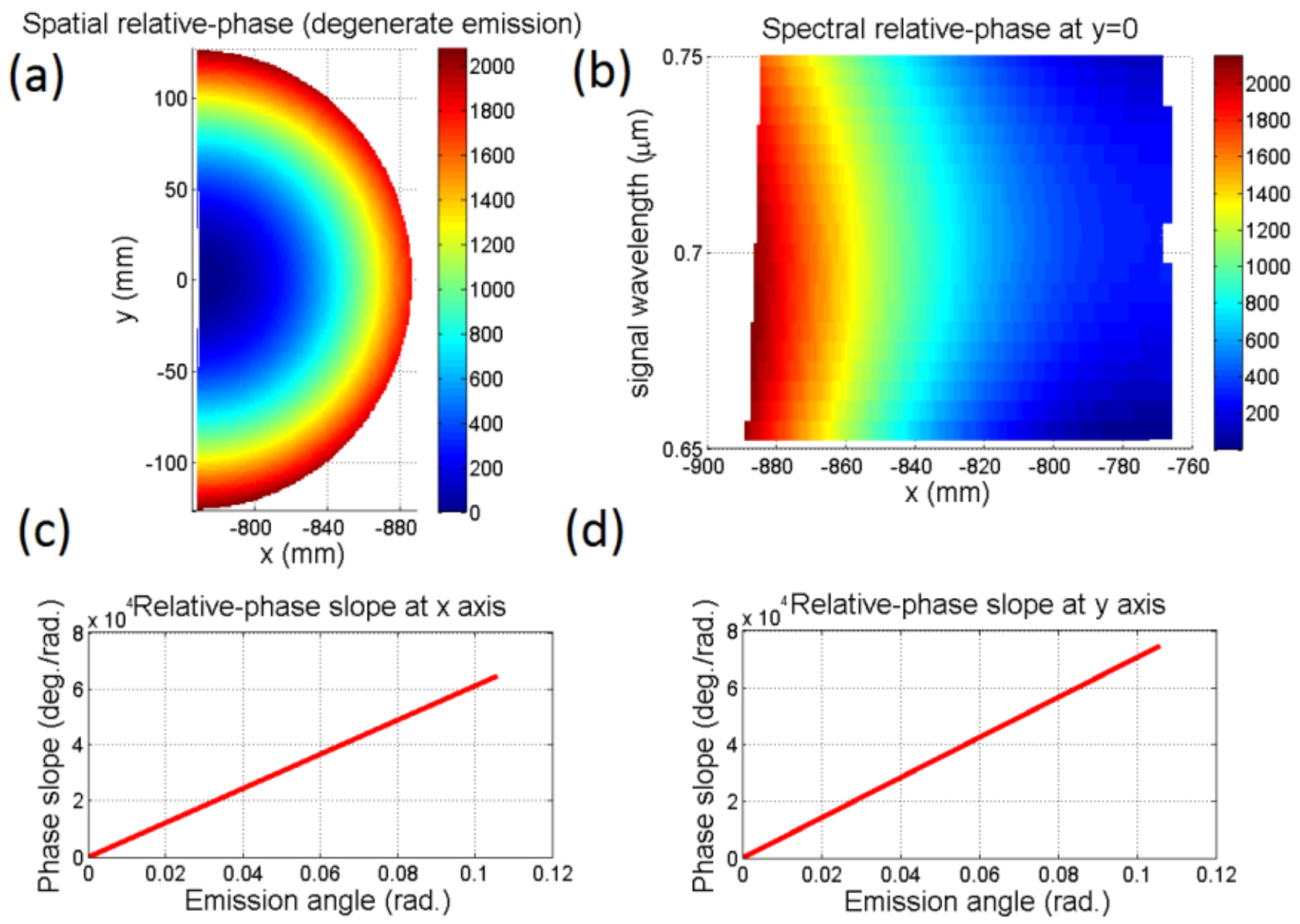

Figure 5. The same as in Fig. 3 for $\theta_p = 20°$ and $\phi_p = 180°$

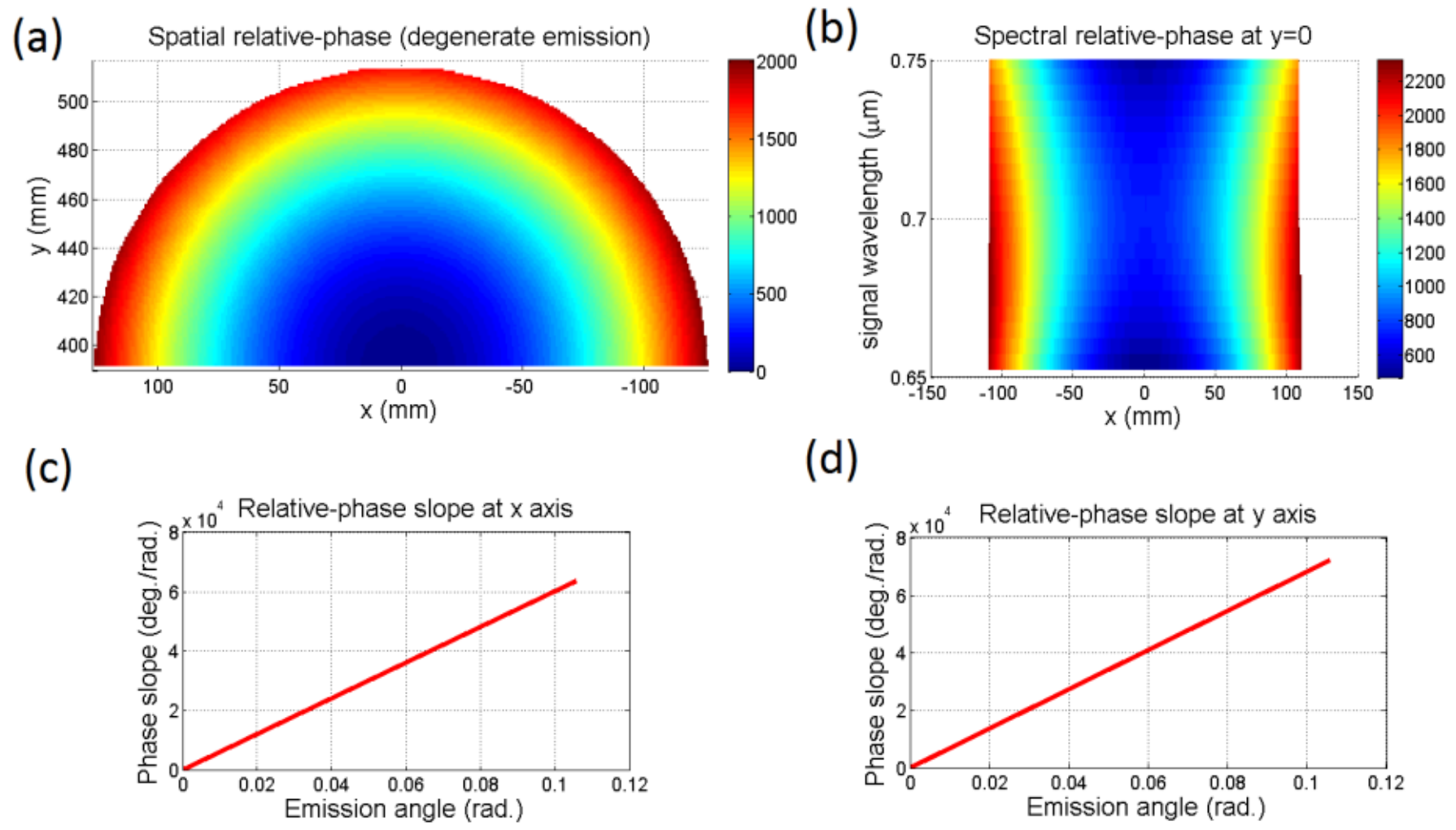

Figure 6. The same as in Fig. 3 for $\theta_p = 10°$ and $\phi_p = 90°$

## 4.2 Time-delay map

Ideally, the superposed wavefunction of the biphoton emerging from the two crystals due to a pump photon shouldn't involve correlations with other domains; e.g., space, time, and frequency. Otherwise, the degree of entanglement is proportionally compromised. The time interval between the instants when the pump photon crosses the free-space : 1$^{st}$-crystal interface and when a photon pair born at a depth $\mu d$ $(\mu \leq 1)$ of the 1$^{st}$ crystal emerges from the 2$^{nd}$-crystal : free-space interface

$$t_{\varepsilon}^{(1)}(w_s, \theta_{sa}, \phi_{sa}) = d\left(\frac{\mu}{V_{pe}} + \frac{1-\mu}{V_{\varepsilon o}K_{\varepsilon o,z}} + \frac{1}{V_{\varepsilon e}r_{\varepsilon,z}}\right), \qquad \varepsilon = s, i \qquad (12)$$

where $V_{pe}$, $V_{\varepsilon e}$ are the group velocities of extraordinarily polarized pump and downconverted photons, $V_{\varepsilon o}$ is the group velocity of ordinarily polarized downconverted photon. The time interval in case if that photon pair is born at equivalent depth in the 2nd crystal

$$t_{\varepsilon}^{(2)}(w_s, \theta_{sa}, \phi_{sa}) = d\left(\frac{1}{V_{po}} + \frac{\mu}{V_{pe}} + \frac{1-\mu}{V_{\varepsilon o}K_{\varepsilon o,z}}\right), \qquad \varepsilon = s, i \qquad (13)$$

The relative time delay can be then calculated as

$$\Delta t_{\varepsilon}(w_s, \theta_{sa}, \phi_{sa}) = d\left(\frac{1}{V_{\varepsilon e}r_{\varepsilon,z}} - \frac{1}{V_{po}}\right), \qquad \varepsilon = s, i \qquad (14)$$

which gains its spectral and angular dependence due only to the first term. The difference between $\Delta t_s$ and $\Delta t_i$ then evolves as we get further from the degenerate phase matching condition[17] and also when the downconversion takes place in a plane around the principal plane of the 2nd crystal. To obtain high-fidelity entangled state, the time delays $\Delta t_{\varepsilon}$ should be kept sufficiently less than the coherence time of the pump photons[11,14,17]. Figures 7-9 show the time delay map for different settings of pump incidence. It is remarkable that in Fig. 9, the pump incidence at $\theta_p = 52°$ and $\phi_p = 90$ deg achieves perfect time-delay compensation in the horizontal plane. This type of self-compensation source can be of special importance in experiments.

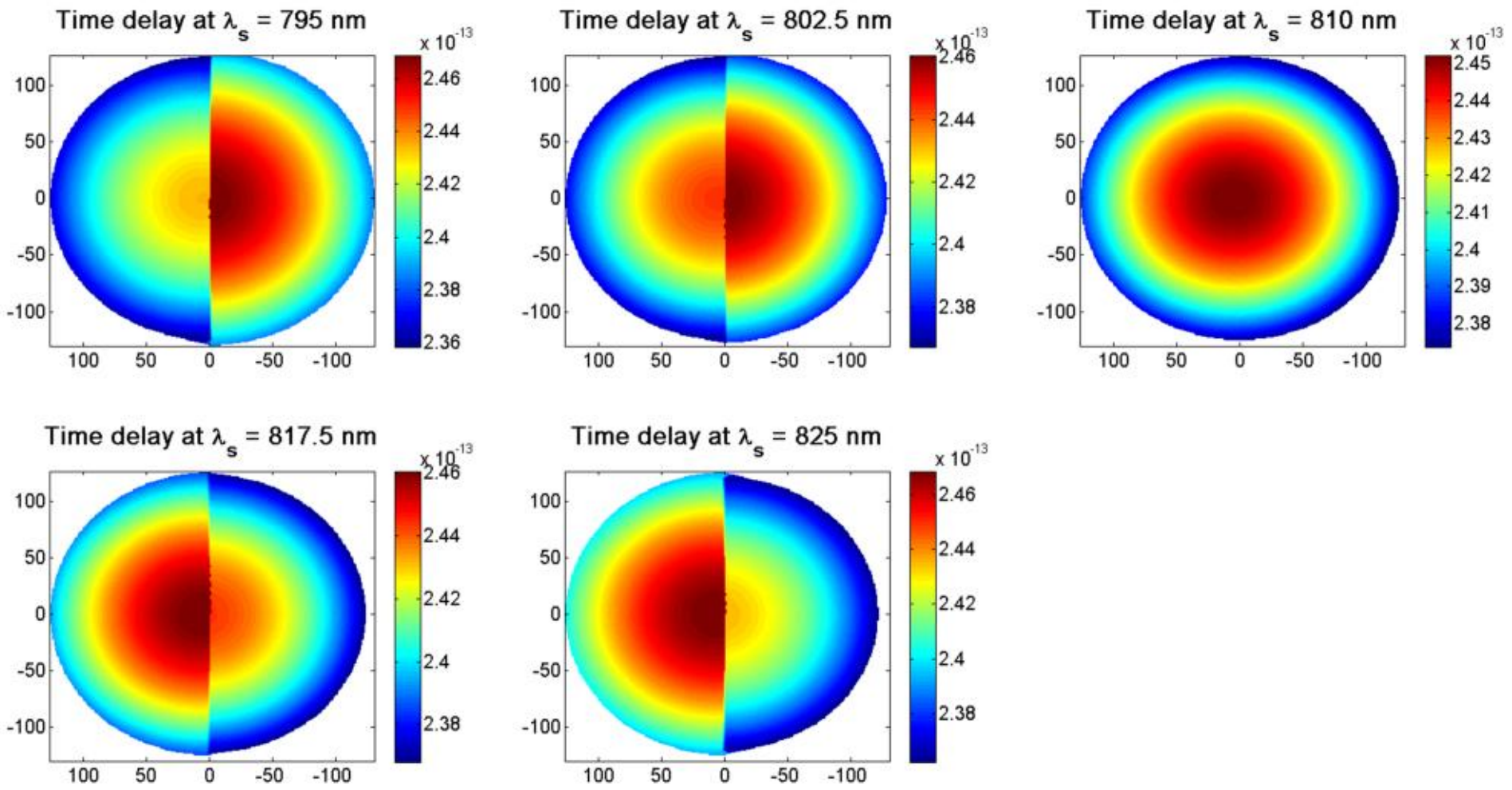

Figure 7. The maps of the time-delay between wavepackets scattered from the 1st crystal and the superimposed ones from the 2nd crystal. To clarify the frequency dependence, we assume the signal photon to be detected behind narrow filter centered at different wavelengths as shown above the subfigures. The left half of each map is reserved for $\Delta t_s$, while $\Delta t_i$ is shown to the right. The downconversion is assumed to take place in two BBO crystals each of length 0.6 mm with the angle between the pump beam and optic axis equals 29.3° The two-crystal are pumped by normally incident beam ($\theta_p = 0°$ and $\phi_p = 0°$) at 405-nm to have degenerate emission centered at ~3°

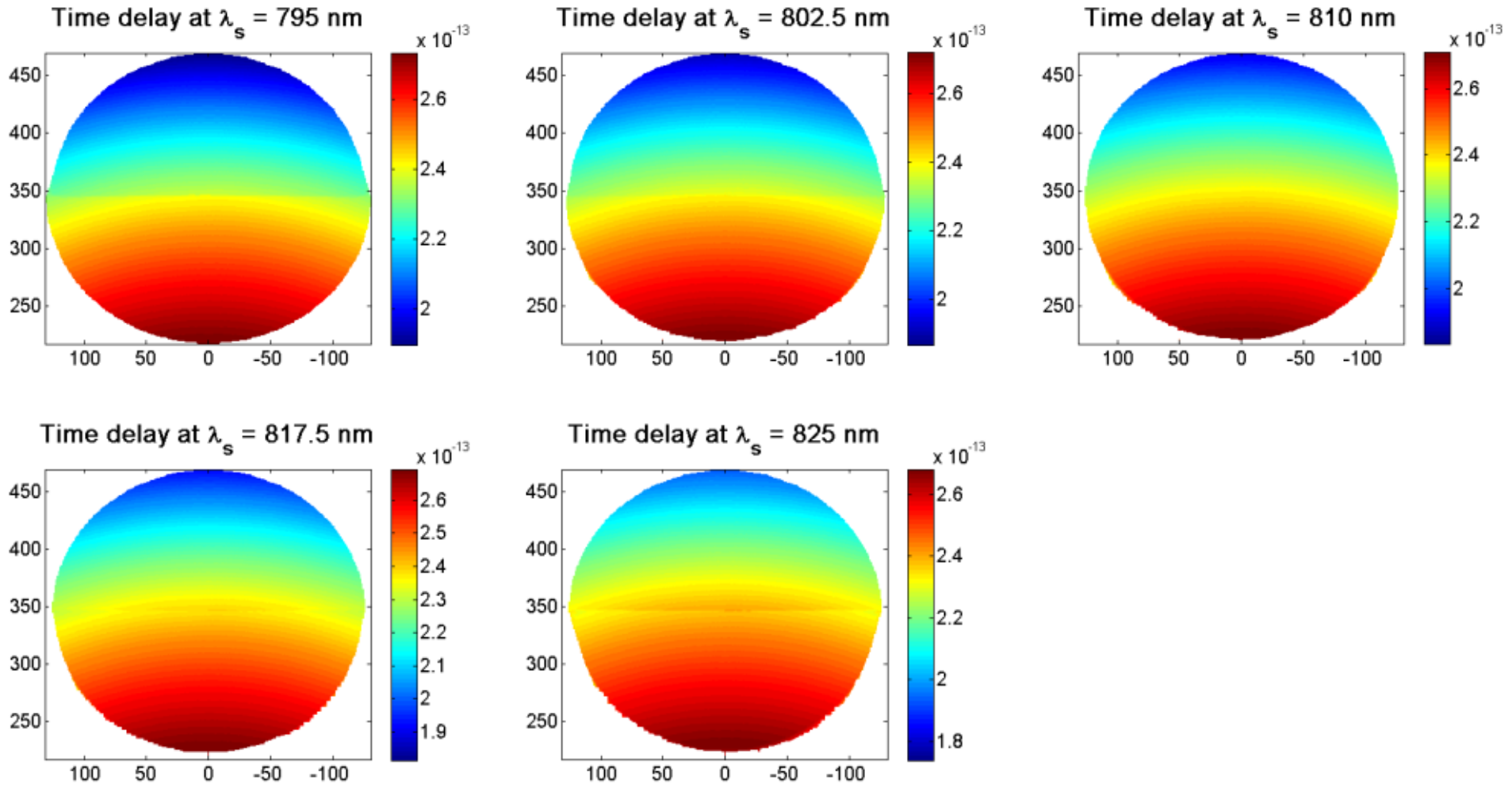

Figure 8. The same as in Fig. 7 for $\theta_p = 10°$ and $\phi_p = 90°$

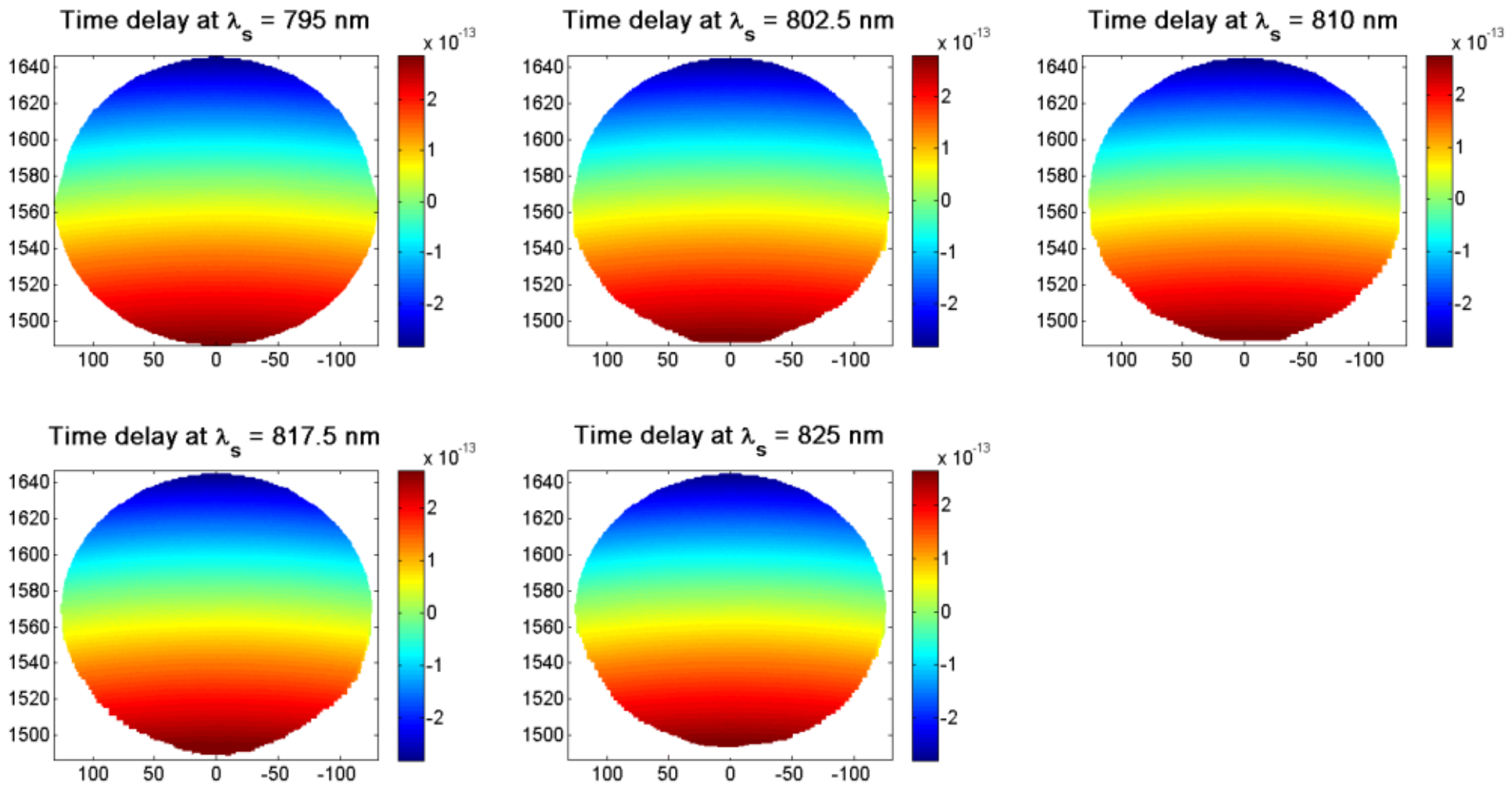

Figure 9. The same as in Fig. 7 for $\theta_p = 52°$ and $\phi_p = 90°$ (time delay compensation in horizontal plane).

## ACKNOWLEDGEMENTS

This work was supported by the Information Technology Industry Development Agency (ITIDA-ITAC), Ministry of Communication, Egypt.

## REFERENCES

[1] Mattle, K. *et al*., "Dense coding in experimental quantum communication," *Phys. Rev. Lett.* **76**, 4656–4659 (1996).

[2] Kim, Y. H., Kulik, S. P. and Shih, Y. *Phys. Rev. Lett.* 86, 1370–1373 (2001).

[3] Kwiat, P. G., and Weinfurter, H., "Embedded Bell-state analysis," *Phys. Rev. A*, 58(4), R2623 (1998).

[4] Schuck, C., Huber, G., Kurtsiefer, C. and Weinfurter, H., "Complete deterministic linear optics Bell state analysis," *Phys. Rev. Lett.* **96**, 190501 (2006).

[5] Barbieri, M., Vallone, G., Mataloni, P. & Martini, F. D., "Complete and deterministic discrimination of polarization Bell states assisted by momentum entanglement," *Phys. Rev. A* **75**, 042317 (2007).

[6] Barreiro, J. T., Wei, T., and Kwiat, P. G., "Beating the channel capacity limit for linear photonic superdense coding," *Nature physics* 4.4 (2008): 282-286.

[7] Graham, T.M., Bernstein, H.J., Wei, T.C., Junge, M. and Kwiat, P.G. "Superdense teleportation using hyperentangled photons," *Nature communications*, *6 (*2015*)*.

[8] Cerf, N. J., Adami, C., and Kwiat, P. G., "Optical simulation of quantum logic," *Phys. Rev. A* 57, R1477–R1480 (1998).

[9] Fiorentino, M. and Wong, F. N. C., "Deterministic controlled-not gate for single-photon two-qubit quantum logic," *Phys. Rev. Lett.* 93, 070502 (2004).

[10] Kwiat, P. G., Waks, E., White, A. G., Appelbaum, I., and Eberhard, P. H., "Ultrabright source of polarization-entangled photons," *Phys. Rev. A* 60, R773–R776 (1999).

[11] Nambu, Y., Usami, K., Tsuda, Y., Matsumoto, K., and Nakamura, K., "Generation of polarization-entangled photon pairs in a cascade of two type-I crystals pumped by femtosecond pulses," *Phys. Rev. A* 66, 033816 (2002).

[12] Altepeter, J. B., Jeffrey, E. R., and Kwiat, P. G., "Phase-compensated ultra-bright source of entangled photons," *Opt. Express* 13, 8951– 8959 (2005).

[13] Akselrod, G. M., Altepeter, J. B., Jeffrey, E. R., and Kwiat, P. G., "Phase-compensated ultra-bright source of entangled photons: erratum," *Opt. Express* 15, 5260–5261 (2007).

[14] Rangarajan, R., Goggin, M., and Kwiat, P. G., "Optimizing type-I polarization-entangled photons," *Opt. Express* 17, 18921–18933 (2009).

[15] Hegazy, S. F., Mansour, M. S., and El-Nadi, L., "Enhanced type-I polarization-entangled photons using CW-diode laser," Proc. MTPR-10 (World Scientific, 2010), pp. 211‑220.

[16] Cialdi, S., Brivio, D., and Paris, M. G. A., "Programmable purification of type-I polarization-entanglement," *Appl. Phys. Lett.* 97, 041108 (2010).

[17] Straupe, S., and Kulik, S., "The problem of preparing entangled pairs of polarization qubits in the frequency-nondegenerate regime," *J. Exp. Theor. Phys.* 110, 185–192 (2010).

[18] Hegazy, S. F., and Obayya, S. S. A., "Tunable spatial-spectral phase compensation of type-I (ooe) hyperentangled photons," *J. Opt. Soc. Am. B*, 32(3), 445-450 (2015).